\documentclass[aps,prd,twocolumn,superscriptaddress,nofootinbib,preprintnumbers]{revtex4-1}
\usepackage{amsmath}
\usepackage{graphicx}
\usepackage{subfigure}
\usepackage{amssymb}
\usepackage{xcolor}
\usepackage{multirow}
\usepackage{cancel}
\usepackage{color}
\usepackage{ulem}
\usepackage{listings}
\usepackage{xcolor}
\usepackage{float}
\usepackage{slashed}
\usepackage{wrapfig}
\usepackage{mathtools}
\usepackage[colorlinks,citecolor=blue]{hyperref}
\usepackage[T1]{fontenc}  
\usepackage[utf8]{inputenc}  
\usepackage{times}
\begin{document}
	
\title{Bounds and detection of  MeV-scale dark matter annihilation to neutrinos}
	
\author{Shinya Kanemura}
\email{kanemu@het.phys.sci.osaka-u.ac.jp}
\affiliation{Department of Physics, The University of Osaka, Toyonaka, Osaka 560-0043, Japan}

\author{Shao-Ping Li}
\email{lisp@het.phys.sci.osaka-u.ac.jp}
\affiliation{Department of Physics, The University of Osaka, Toyonaka, Osaka 560-0043, Japan}
	
\author{Dibyendu Nanda}
\email{dnanda@het.phys.sci.osaka-u.ac.jp}
\affiliation{Department of Physics, The University of Osaka, Toyonaka, Osaka 560-0043, Japan}

\begin{abstract}
Current and  most upcoming neutrino detectors can only reach  a  dark matter annihilation cross section to neutrinos  larger than the standard  freeze-out value, but they open intriguing detection avenues  for non-standard  dark matter paradigms. An important corollary  of these non-standard scenarios is  relic  dark matter annihilation after neutrino decoupling, which  was previously overlooked in constraining   MeV-scale dark matter. However,  by combining the  contributions from entropy injection  during neutrino decoupling and from  nonthermal neutrino energy release after decoupling, we  derive significant constraints on the annihilation cross section to neutrinos, which in some mass regimes become stronger than the current bounds. Furthermore, we find that the lower bounds on dark matter masses become inconclusive under the recent data releases from the DESI, SPT-3G, and ACT collaborations. These bounds determine the extent to which upcoming neutrino detectors will probe dark matter annihilation into neutrinos.
\end{abstract}

\maketitle

 \preprint{OU-HET 1273} 
	
\section{Introduction}	\label{sec:intro}
Dark matter (DM) annihilation into the Standard Model (SM) particles  represents one of the primary avenues for indirect DM detection. Currently, DM annihilation into electromagnetic species is strongly constrained by present-day astrophysical and late-time cosmological  observations, as well as by observables inherited from the early Universe - including  diffuse photon background,   gamma-ray,  cosmic microwave background (CMB) anisotropies, CMB spectral distortions and big-bang nucleosynthesis (BBN). In contrast, DM annihilation into neutrinos remains the least constrained, due to the weakly interacting nature of neutrinos, which makes them effectively invisible to standard detection methods. This feature can naturally account for the continued non-observation of DM-SM interactions.  Despite the challenges of  neutrino detection, it offers invaluable opportunities to probe various DM particles that exclusively annihilate into neutrinos.

 MeV-scale DM annihilating into neutrinos is a promising target for several low-energy neutrino experiments, including Borexino~\cite{Borexino:2010zht,Borexino:2019wln}, KamLAND~\cite{KamLAND:2011bnd,KamLAND:2021gvi}, Super-Kamiokande (SK)~\cite{Super-Kamiokande:2005wtt,Super-Kamiokande:2008ecj,Super-Kamiokande:2010tar,Super-Kamiokande:2011lwo,Olivares-DelCampo:2017feq}, Hyper-Kamiokande (HK)~\cite{Olivares-DelCampo:2018pdl,Bell:2020rkw}, and JUNO~\cite{JUNO:2015zny,Akita:2022lit,JUNO:2023vyz}. These detectors can search for neutrino fluxes originating from galactic/extragalactic sources, such as DM annihilation and diffuse supernova neutrino background. However, the current, and even the proposed, sensitivities of these experiments can hardly  reach the annihilation cross section required to produce the correct DM relic abundance,  $\langle \sigma v\rangle_{\rm th}\simeq 3\times 10^{-26}\text{cm}^{3}/\text{s}$ ~\cite{Steigman:2012nb,Arguelles:2019ouk,Dutta:2022wdi,Chu:2023jyb}, as  predicted  by the standard thermal DM freeze-out paradigm~\cite{Kolb:1990vq}, and hence cannot decisively test the standard scenario. Nevertheless, this limitation can be turned into an advantage, as these experiments provide invaluable avenues for probing more general DM scenarios, where the annihilation cross section into neutrinos is larger while still consistent with the observed relic density. In fact, neutrino detection may become the only viable way to uncover the origin and particle nature of DM when it interacts with SM particles solely via the neutrino portal~\cite{Batell:2017cmf,Ballett:2019pyw,Blennow:2019fhy,Li:2022bpp,Xu:2023xva}.
 
 There exists a broad class of DM models where annihilation dominantly occurs into SM neutrinos, often motivated by connections to the origin of light neutrino masses~\cite{Tao:1996vb,Ma:2006km,Batell:2017cmf,Ballett:2019pyw,Blennow:2019fhy,Li:2022bpp,Xu:2023xva}, among many other possibilities. In such scenarios, DM can establish thermal contact with neutrinos, even though the relic abundance is not  determined by freeze-out via annihilation into neutrinos. For $\langle \sigma v\rangle > \langle \sigma v\rangle_{\rm th}$, a regime accessible to current and future neutrino detectors, an important consequence of these non-standard scenarios is that significant DM annihilation may occur well after the neutrino decoupling epoch. This late-time annihilation has not been taken into account in conventional constraints on MeV-scale DM masses~\cite{Palomares-Ruiz:2007trf,Ho:2012ug,Boehm:2013jpa,Nollett:2014lwa,Escudero:2018mvt,Sabti:2019mhn,EscuderoAbenza:2020cmq,Sabti:2021reh,Chu:2022xuh}. It is important to note that the lower mass bounds derived from the effective number of neutrino species, $N_{\rm eff}$, are based on the evolution of the neutrino temperature assuming thermal distribution functions for both neutrinos and DM. However, once the neutrino decoupling process is complete, further DM annihilation does not alter the thermal neutrino temperature. Instead, the neutrinos produced from relic DM annihilation contribute directly to the nonthermal neutrino energy density. This effect is analogous to DM annihilation into photons after the Universe cools below 1~keV, when the energy injection can no longer be described by a fully thermal distribution~\cite{Chluba:2013wsa,Chluba:2013pya,Li:2024xlr}.

In this work, we elaborate on the effects of relic DM annihilation into neutrinos after the completion of neutrino decoupling, and derive the  bounds on the constant (DM velocity independent)  annihilation cross section as a function of the DM mass, while accounting for entropy production during the neutrino decoupling era. To this end, we utilize the latest measurements of $N_{\rm eff}$, from the Dark Energy Spectroscopic Instrument (DESI) Data Release 1 and 2~\cite{DESI:2024mwx,DESI:2025ejh}, South Pole Telescope (SPT-3G)~\cite{SPT-3G:2024atg}, Atacama Cosmology Telescope (ACT) Data Release 6~\cite{ACT:2025tim}, and Planck 2018 ~\cite{Planck:2018vyg}. We find that the resulting lower bounds on the DM mass are sensitive to the updated measurements of $N_{\rm eff}$ and exhibit noticeable variation across datasets. As a result, a definitive lower limit on the thermal DM mass has yet to be established.

Depending on the datasets used, the resulting DM mass bounds can significantly impact the extent to which upcoming neutrino detectors, such as JUNO with 20 years of data, can probe MeV-scale DM annihilation into neutrinos. We illustrate this by highlighting the interplay between the detection sensitivity to $\langle \sigma v \rangle$ and modifications to $N_{\rm eff}$ in the range of $0.2$–$0.4$, which is favored by several studies aiming to alleviate the Hubble tension~\cite{DiValentino:2021izs,Schoneberg:2021qvd,Kamionkowski:2022pkx,DiValentino:2025sru}. We also find that the constraints arising from relic DM annihilation are comparable to current bounds obtained from observations of astrophysical neutrino fluxes. For DM masses approaching the lower limits, these cosmological constraints   even surpass those from the current detectors.

\section{Thermal MeV DM with relic annihilation}	\label{sec:MeV-DM}
When MeV-scale DM has an averaged constant annihilation cross section larger than $\langle \sigma v\rangle_{\rm th}$, it generally keeps DM in thermal equilibrium with neutrinos  throughout the entire epoch of neutrino decoupling. In such a case, entropy injection by  DM annihilation modifies the evolution of  neutrino and photon temperatures, resulting in a temperature ratio $T_\nu/T_\gamma$ that deviates from the SM prediction at the time of neutrino decoupling. This, in turn, leads to a prediction of a larger  $N_{\rm eff}$. Precise measurements of the CMB and BBN have  placed lower  bounds on the DM mass, typically in the range of 1 MeV to 10 MeV, for thermal DM candidates ~\cite{Ho:2012ug,Boehm:2013jpa,Nollett:2014lwa,Escudero:2018mvt,Sabti:2019mhn,EscuderoAbenza:2020cmq,Sabti:2021reh,Chu:2022xuh}.

The lower mass bounds depend primarily on the spin of the DM particle and are largely independent of the annihilation cross section. This is because the evolution of neutrino and photon temperatures during the non-instantaneous neutrino decoupling epoch ($0.01~\text{MeV} \lesssim T \lesssim 1~\text{MeV}$) is influenced by entropy injection arising from the thermal DM transition from the relativistic to the nonrelativistic regime. While the annihilation cross section is often fixed at the thermal value, $\langle \sigma v\rangle = \langle \sigma v\rangle_{\rm th}$ ~\cite{Escudero:2018mvt,Sabti:2019mhn,EscuderoAbenza:2020cmq}, a larger cross section still ensures thermalization between DM and neutrinos and does not cause significant differences from the case of $\langle \sigma v\rangle=\langle \sigma v\rangle_{\rm th}$, since  the DM freeze-out temperature depends only logarithmically on the annihilation cross section~\cite{Kolb:1990vq}. As a result, the lower mass bounds remain applicable even for   larger cross sections that could be probed by the neutrino detectors. Nevertheless, an annihilation cross section to neutrinos larger than $\langle \sigma v\rangle_{\rm th}$ points to a non-standard  DM production mechanism, where   late-time DM creation after freeze-out must be present  to    account for the enhanced depletion through the annihilation.  

 Naively, one might envisage that late-time annihilation, occurring after DM and neutrino decoupling, would have a small effect on $N_{\rm eff}$, since the DM number density is significantly reduced compared to its   equilibrium density.  Nevertheless, there are three  factors  that make the late-time modification nontrivial. Firstly, entropy injection during neutrino decoupling is shared between the neutrino and photon sectors, especially via electron-positron annihilation. This implies that even if DM exclusively annihilates into neutrinos, the resulting increase in neutrino temperature can still influence the photon temperature through the coupled Boltzmann equations. After neutrino decoupling, however, DM annihilation into neutrinos primarily contributes to an additional accumulation of neutrino energy over time.
 
The second factor can be seen by the definition of $N_{\rm eff}$ from nonthermal energy release 
\begin{align}
 N_{\rm eff}^{\rm nth}=\frac{\rho_{\nu}^{\rm nth}}{\rho_\nu^{\rm SM}}\,,
\end{align}
where $\rho_{\nu}^{\rm SM}$  denotes the one-flavor SM neutrino energy density
\begin{align}
\rho_{\nu}^{\rm SM}=\frac{7\pi^2}{120}T_\nu^4\,,
\end{align}
 with the neutrino-photon temperature ratio $T_\nu/T_{\gamma}=(4/11)^{1/3}$. Right  after the neutrino energy release, both $\rho_{\nu}^{\rm SM}$ and $\rho_{\nu}^{\rm nth}$ scale identically as the Universe expands, such that  $N_{\rm eff}^{\rm nth}$ becomes constant. While later  neutrino energy release  corresponds to a lower DM energy density and hence smaller $\rho_{\nu}^{\rm nth}$,   the background neutrino energy density also redshifts to lower values, implying  that  late-time energy release may not be negligibly small. 
 
The third and more important factor arises from the combined effect of entropy injection and energy release. If DM remains in thermal equilibrium with neutrinos before decoupling is complete, it already contributes to $N_{\rm eff}$  through entropy injection. This early modification of $N_{\rm eff}$ leaves less room for late-time relic DM annihilation to have a significant impact, implying that even a suppressed late-time energy release could still produce observable effects. Therefore, it is worth carefully investigating whether relic DM annihilation after DM/neutrino decoupling can contribute to $N_{\rm eff}$  in a non-negligible way.

The modification of  $N_{\rm eff}$ can be parameterized as
\begin{align}
	N_{\rm eff}\equiv \frac{8}{7}\left(\frac{11}{4}\right)^{4/3}\left(\frac{\rho_{\nu}^{\rm th}+\rho_{\nu}^{ \rm nth}}{\rho_\gamma}\right),
\end{align}
where $\rho_{\nu}^{\rm th}$ denotes the thermal  neutrino energy density, and $\rho_{\nu}^{\rm nth}$ the nonthermal one produced from relic DM annihilation after neutrino decoupling is well complete.  Using $\rho_\gamma=\pi^2T^4_\gamma/15$,  we can rewrite the thermal part as
\begin{align}\label{eq:Neff-eq}
	N_{\rm eff}^{\rm th}=3\left(\frac{11}{4}\right)^{4/3}\left(\frac{T_\nu}{T_\gamma}\right)^4\,.
\end{align}
Following Ref.~\cite{Escudero:2018mvt}, one sees that $T_\gamma/T_\nu\approx 1.3958$   corresponds to $N_{\rm eff}=3.045$ as the prediction in the SM. Including the thermal DM energy and entropy to the evolution of the neutrino and photon temperatures will eventually yield a larger  $T_\nu/T_\gamma$ and hence a larger $N_{\rm eff}^{\rm th}$. To this end, we use the approach presented in Ref.~\cite{Escudero:2018mvt}, where the reduced Boltzmann equations for $T_\nu, T_\gamma$ read
\begin{align}\label{eq:dTnudt}
	\frac{dT_\nu}{dt}&=\frac{-12H \rho_{\nu}+d\rho_{\rm DM}/dt+\delta \rho_\nu/\delta t}{3\partial \rho_{\nu}/\partial T_\nu+\partial \rho_{\rm DM}/\partial T_\nu}\,,
	\\[0.2cm]
		\frac{dT_\gamma}{dt}&=\frac{-4H \rho_{\gamma}+d\rho_e/dt-P_{\rm int}^{(1)}-\delta \rho_\nu/\delta t}{\partial \rho_{\gamma}/\partial T_\gamma+\partial \rho_{e}/\partial T_\gamma+P_{\rm int}^{(2)}}\,,\label{eq:dTgammadt}
\end{align}
where 
\begin{align}
\frac{d\rho_{\rm DM}}{dt}&=-3H(\rho_{\rm DM}+p_{\rm DM})\,, 
\\[0.2cm]
\frac{d\rho_e}{dt}&=-3H(\rho_{e}+p_{e})\,,
\end{align}
with $\rho_i, p_i$ the thermal energy density and pressure. We have defined 
\begin{align}
 P_{\rm int}^{(1)}\equiv -3H T_\gamma \frac{dP_{\rm int}}{dT_\gamma}\,,\quad
P_{\rm int}^{(2)}\equiv T_\gamma \frac{d^2 P_{\rm int}}{dT_\gamma^2}\,, 
 \end{align}
 as the finite-temperature corrections, where the  first-order correction of the electromagnetic pressure   $P_{\rm int}$ can be  found  in Ref.~\cite{Mangano:2001iu}. The   energy release rate $\delta \rho_\nu/\delta t$ denotes contributions  from electron-positron annihilation and electron-neutrino (neutrino-neutrino) scattering, which is shared between the neutrino and photon temperature evolution due to energy conservation. This makes the two Boltzmann equations coupled, such that a new-physics effect on increasing $T_\nu$ will also modify $T_\gamma$. The detailed expressions of $\delta \rho_\nu/\delta t$ are known and can be found e.g. in Ref.~\cite{Hannestad:1995rs,Escudero:2018mvt}.

The  coupled Boltzmann equations Eqs.~\eqref{eq:dTnudt}-\eqref{eq:dTgammadt} assume fully thermal contact between DM and neutrinos sharing a common temperature, where DM-related dynamics depends only on the DM mass and its spin.  Due to such dependence,    measurements of $N_{\rm eff}$ from BBN and CMB can set constraints on the DM mass and spin.  Note that, the thermally averaged annihilation cross section does not essentially enter the Boltzmann equations provided that it is large enough to maintain   thermal equilibrium at least after electron-positron annihilation. For an even larger annihilation cross section, the DM freeze-out temperature is lower, whose dependence on $\langle \sigma v\rangle$ is however only logarithmic~\cite{Kolb:1990vq}. For DM masses above 1~MeV,  DM thermal freeze-out via annihilation to neutrinos will typically  be around the moment when the $T_\nu/T_\gamma$  evolution freezes. 

After DM and neutrinos decouple from the thermal plasma, both Eq.~\eqref{eq:dTnudt} and Eq.~\eqref{eq:dTgammadt} cannot be applied. Instead, the injected neutrinos from relic DM annihilation directly contribute to extra nonthermal neutrino energy density, where 
the modification of $N_{\rm eff}$ reads
\begin{align}\label{eq:Neff-neq}
	N_{\rm eff}^{\rm nth}=\frac{8}{7}\left(\frac{11}{4}\right)^{4/3}\left(\frac{\rho_{\nu}^{ \rm nth}}{{\rho_\gamma}}\right).
	\end{align}
The treatment for determining such relic contributions is   easier than those contributions from solving the coupled Boltzmann equations. 
The collision rate of nonthermal neutrino energy release   after the completion of neutrino decoupling gives 
 \begin{align}
\mathcal{C}_\nu&=\langle \sigma v\rangle m_{\rm DM}n_{\rm DM}^2
	 \\[0.2cm]
&\approx  c \left(\frac{T}{0.1~\text{MeV}}\right)^6\left(\frac{\langle \sigma v\rangle }{\langle\sigma v\rangle_{\rm th}}\right)\left(\frac{1~\text{MeV}}{m_{\rm DM}}\right)\text{MeV}^5\,,\label{eq:s-wave-rate}
\end{align} 
where $c=1.32\times 10^{-33}$ and  we have fixed a factor   $(\Omega_{\rm DM}h^2/0.12)^2=1$ from the $n_{\rm DM}^2$ dependence. Note that we have also assumed relic DM annihilation occurs in the nonrelativistic   regime. 

Defining $\rho_{\nu}^{\rm nth}\equiv Y_\nu s_{\rm SM}^{4/3}$, we can rewrite 
the  Boltzmann evolution of the total neutrino and antineutrino energy release as
\begin{align}\label{eq:dYdT}
\frac{dY_\nu}{dT}=-2\frac{\mathcal{C}_\nu}{s_{\rm SM}^{4/3}H T}\,,
\end{align}
where we set $T\equiv T_\gamma$, and the factor of 2 accounts for energy densities from neutrinos and antineutrinos.  
The  Hubble parameter at radiation-dominated epoch reads
\begin{align}
H\approx 1.66\sqrt{g_\rho(T)}\frac{T^2}{M_{\rm Pl}}\,,
\end{align}
 with  $M_{\rm P}\approx 1.22\times 10^{19}$~GeV the Planck mass, and
 \begin{align}
  s_{\rm SM}=\frac{2\pi^2}{45}g_{s}(T)  T^3
  \end{align}
  denotes the SM entropy density. In the following, we will set $g_\rho\approx g_{s}\approx 3.34$ for the relativistic degrees of freedom in energy and entropy densities after the completion of neutrino decoupling, which is justified as a good approximation after the QCD phase transition~\cite{Borsanyi:2016ksw}.
  
\begin{figure*}[t]
	\centering
	\includegraphics[scale=0.33]{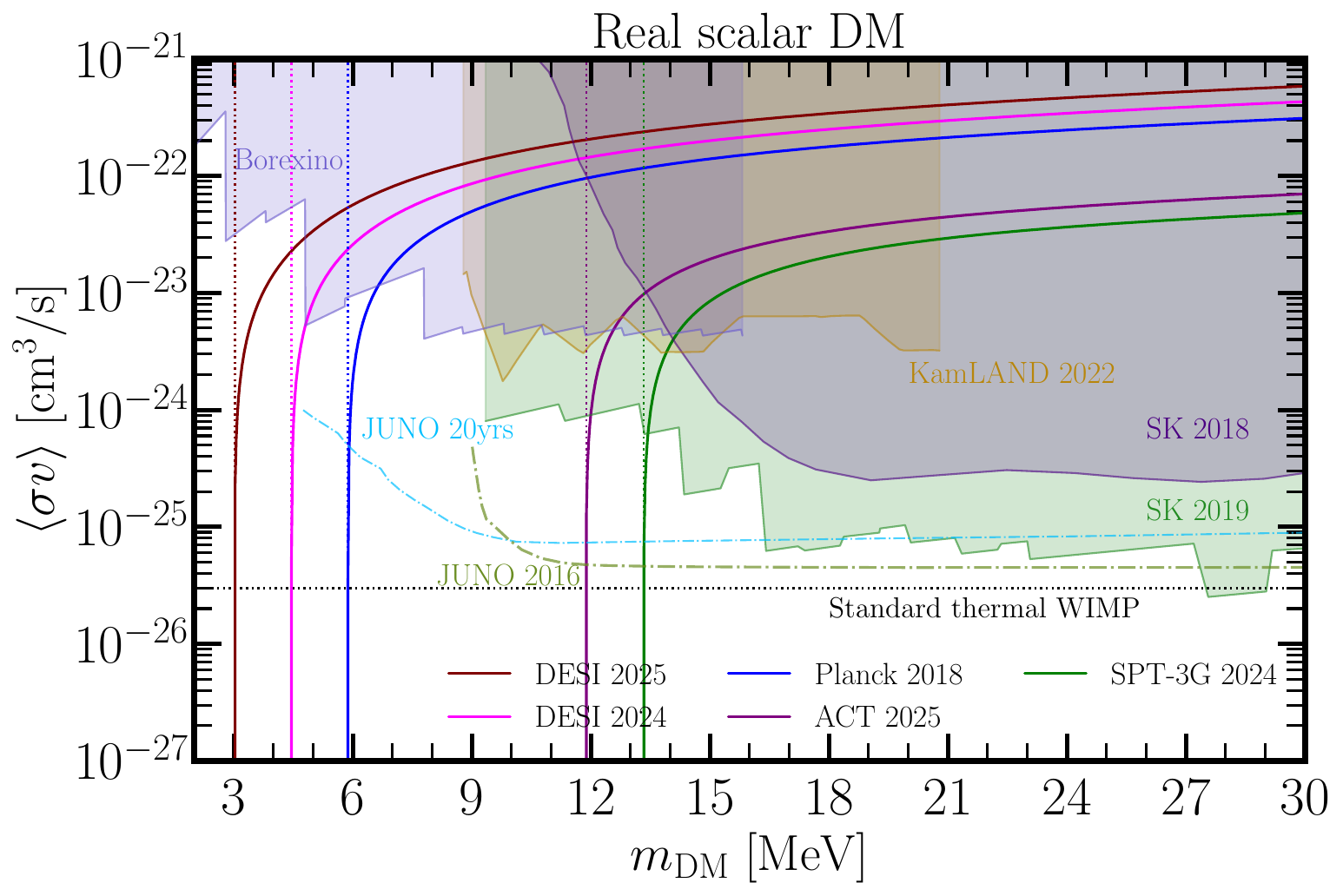} \quad \includegraphics[scale=0.33]{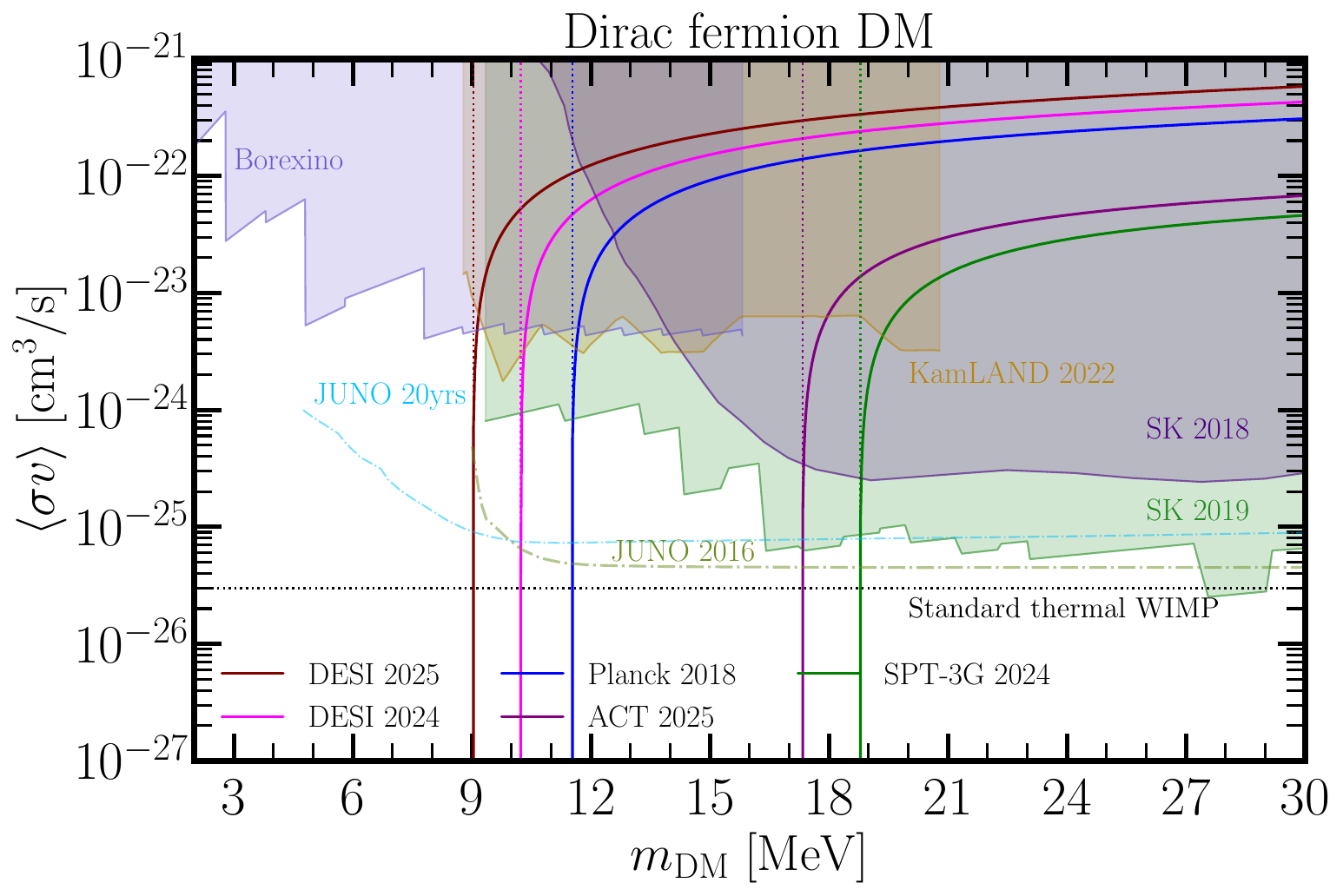}
	\caption{\label{fig:th-bound}$N_{\rm eff}$ bounds on thermal scalar DM (left panel) and Dirac fermion DM (right panel)  with relic annihilation to neutrinos. Shaded regions represent the current bounds from neutrino flux observations in neutrino detectors, while dashed-dotted lines denote the forecast sensitivities of upcoming JUNO experiments. See the text for more details. }
\end{figure*}

It is worthwhile  to mention that  the right-hand side of Eq.~\eqref{eq:dYdT}   scales as  $1/T$ for a constant cross section $\langle \sigma v\rangle$, indicating that  the nonthermal neutrino energy release is   dominated at lower temperatures, and hence the relic DM annihilation cannot be neglected. In addition,  the resulting  $Y_\nu$ has a  logarithmic dependence on the temperature. Indeed, we found 
 \begin{align}\label{eq:Neff-neq-2}
N_{\rm eff}^{\rm nth}=0.002\left(\frac{\langle\sigma v\rangle}{10^{-24}}\right) \left(\frac{1~\text{MeV}}{m_{\rm DM}}\right) \ln\left(\frac{T_{\rm dec}}{T_{\rm CMB}}\right),
 \end{align}
 where $T_{\rm dec}=\mathcal{O}(0.01)$~MeV denotes the moment for the completion of non-instantaneous neutrino decoupling, and $T_{\rm CMB}=\mathcal{O}(0.1)$~eV the recombination epoch. Varying $T_{\rm dec}, T_{\rm CMB}$ by an order of magnitude generally only causes a difference of $N_{\rm eff}^{\rm nth}$ at $\mathcal{O}(0.1)\%$. Typically we have  $\ln(T_{\rm dec}/T_{\rm CMB})\approx 12$, and hence the contribution to $N_{\rm eff}$ from   relic DM annihilation is generally larger than $3\%$, a value that can be probed by future CMB experiments, such as the CMB-S4~\cite{CMB-S4:2016ple,Abazajian:2019eic} and CMB-HD~\cite{CMB-HD:2022bsz}.

We combine the contributions from Eq.~\eqref{eq:Neff-eq} and Eq.~\eqref{eq:Neff-neq} to compute the total $N_{\rm eff}$. For  $N_{\rm eff}^{\rm th}$, we set the   evolution start from a high temperature, $T=10$~MeV, where neutrinos, photons and DM are in thermal equilibrium, down to a low temperature $T=10$~keV, when the temperature ratio $T_\nu/T_\gamma$ essentially becomes constant~\cite{Escudero:2018mvt,EscuderoAbenza:2020cmq}. For  $N_{\rm eff}^{\rm nth}$, we set $T_{\rm dec}=10$~keV and $T_{\rm CMB}=0.1$~eV. We show the results in  the left panel of Fig.~\ref{fig:th-bound} for  real scalar DM with one internal  degree of freedom (dof) and in the  right panel for Dirac fermion DM with four internal dof. The vertical dotted lines, corresponding to different datasets taken into consideration, represent the lower mass bounds derived from $N_{\rm eff}^{\rm th}$ while the   $m_{\rm DM}$-$\langle \sigma v \rangle$ curves denote the bounds caused by   $N_{\rm eff}^{\rm nth}$.

We see that the lower mass bounds depend sensitively on what data is taken into account. For Planck 2018, we take
\begin{align}
N_{\rm eff}=2.99^{+0.34}_{-0.33}\,, \quad \text{Planck 2018}
\end{align}
 at $95\%$ confidence level from CMB anisotropies combined with lensing and baryon acoustic oscillation (BAO) measurements~\cite{Planck:2018vyg}. The resulting bounds are 5.9~MeV for real scalar and  11.5~MeV for Dirac  fermion DM, as similarly obtained earlier in Ref.~\cite{Sabti:2019mhn}. Note that the bounds from Majorana fermion DM (dof=2), complex scalar (dof=2) and  vector boson (dof=3) DM will reside in the regions between the real scalar and Dirac fermion cases.
 
For the bounds derived from  the  first year of observations
in DESI Data Release 1, we take the results obtained by combing the CMB anisotropies and DESI BAO~\cite{DESI:2024mwx}:
\begin{align}
	N_{\rm eff}=3.10\pm 0.17\,,\quad \text{DESI 2024}
\end{align}
which, bearing a $2\sigma$ level of uncertainty, yields  weaker lower bounds on the DM mass. For real scalar DM, the lower mass bound is   4.5~MeV, and  for Dirac fermion DM, we obtain $m_{\rm DM}>10.2$~MeV. 

The favored $N_{\rm eff}$ value is slightly larger after the DESI Data Release 2~\cite{DESI:2025ejh},
\begin{align}
	N_{\rm eff}=3.23^{+0.35}_{-0.34}\,, \quad \text{DESI 2025}
\end{align}
at $95\%$ confidence level by combining DESI BAO and CMB anisotropies. This result will allow a shift of $N_{\rm eff}$: $\Delta N_{\rm eff}=0.58$, rendering the bounds of the DM mass even weaker. We find that the lower mass bound is   3~MeV for real scalar DM, and  9~MeV for Dirac fermion DM.

With the Data Release 6 from the ACT collaboration very recently, we apply the data combined with astrophysical measurements of primordial helium and deuterium abundances~\cite{ACT:2025tim}:
\begin{align}
	N_{\rm eff}=2.89\pm 0.11\,,\quad \text{ACT 2025}
\end{align}
which, again bearing a $2\sigma$ level of uncertainty, yields  $m_{\rm DM}>11.9$~MeV for real scalar DM and  $m_{\rm DM}>17.3$~MeV for Dirac fermion DM.

Finally, the results from SPT-3G represent the strongest lower mass bounds. Adopting the data by combining SPT, Planck,  BAO and earlier ACT measurements~\cite{SPT-3G:2024atg}:
\begin{align}
N_{\rm eff}=2.83\pm 0.13\,, \quad \text{SPT-3G 2024}
\end{align}
we find   $m_{\rm DM}>13.3$~MeV for real scalar DM and  $m_{\rm DM}>18.8$~MeV for Dirac fermion   DM. It is worthwhile to emphasize that     with this $N_{\rm eff}$ result, the present-day Hubble parameter $H_0$  shows a $5.4\sigma$ tension with the latest SH0ES measurement~\cite{Breuval:2024lsv}. Therefore, while   SPT-3G represents the strongest lower bounds on DM masses, it may still be premature to make these bounds conclusive, particularly when relaxing  the Hubble tension favors a  larger $N_{\rm eff}$.

The bounds from BBN observations are generally weaker than from combining CMB anisotropies and BAO~\cite{Sabti:2019mhn,Fields:2019pfx,Sabti:2021reh,Yeh:2022heq}. For example, combining BBN with the measurements of primordial ${}^4\text{He}$ and deuterium abundances  still  allows a $95.45\%$ upper limit $\Delta N_{\rm eff}=0.407$~\cite{Yeh:2022heq}. 

Constraints from neutrino detectors shown in shaded regions of Fig.~\ref{fig:th-bound} include the  Borexino collaboration~\cite{Borexino:2010zht,Borexino:2019wln} where   upper limits of neutrino flux  was used in Ref.~\cite{Arguelles:2019ouk}, the updated measurements from KamLAND   where the $J$-factor of the angular-averaged intensity over the whole Milky Way is taken to be 1.3~\cite{KamLAND:2021gvi} (KamLAND 2022), data of three
different SK phases~\cite{Super-Kamiokande:2005wtt,Super-Kamiokande:2008ecj,Super-Kamiokande:2010tar,Super-Kamiokande:2011lwo} used in Ref.~\cite{Olivares-DelCampo:2017feq} (SK 2018), and data from SK phase IV~\cite{Super-Kamiokande:2013ufi} performed in Ref.~\cite{Arguelles:2019ouk} (SK 2019). The forecast sensitivity from JUNO 2016 was performed in  Ref.~\cite{Arguelles:2019ouk} based on the background estimates for diffuse supernova background searches~\cite{JUNO:2015zny}, while that from JUNO 20yrs was performed in Ref.~\cite{Akita:2022lit} with 20 years of   data-taking under  the standard Navarro-Frenk-White DM profile. The standard thermal WIMP case shown in the black dotted line corresponds to $\langle \sigma v\rangle_{\rm th}\simeq 3\times 10^{-26}\text{cm}^{3}/\text{s}$. See Refs.~\cite{Steigman:2012nb,Chu:2023jyb} for more precise determination of $\langle \sigma v\rangle_{\rm th}$.

We see from Fig.~\ref{fig:th-bound} that the varying lower mass bounds significantly affects the extent to which the upcoming neutrino detection can probe MeV-scale thermal DM with an annihilation cross section larger than $\langle\sigma v\rangle_{\rm th}$. As seen in Fig.~\ref{fig:th-bound}, with the severe bounds from SK~\cite{Olivares-DelCampo:2017feq,Arguelles:2019ouk}, there is still large parameter space that will be targeted by JUNO for $m_{\rm DM}\lesssim 15$~MeV.

While not explicitly shown in Fig.~\ref{fig:th-bound}, it is worth mentioning that  there is also large parameter space that can be probed with the forecast HK sensitivity~\cite{Horiuchi:2008jz,Hyper-Kamiokande:2018ofw,Bell:2020rkw} for $10^{-25}\text{cm}^3/\text{s}\lesssim \langle\sigma v\rangle<10^{-24}~\text{cm}^3/\text{s}$ and $m_{\rm DM}>10$~MeV. DM with mass  above 10~MeV up to 1~GeV   may also be covered by future tonne-scale DM direct detection experiments such as DARWIN and ARGO~\cite{McKeen:2018pbb}, though they are under planning phases. With the increased sensitivity from DUNE~\cite{DUNE:2015lol,Capozzi:2018dat}, a DM mass above 100~MeV and the annihilation cross section down to the level of  $10^{-24}~\text{cm}^3/\text{s}$ can be reached~\cite{Arguelles:2019ouk}. 

\begin{figure}[t]
	\centering
	\includegraphics[scale=0.33]{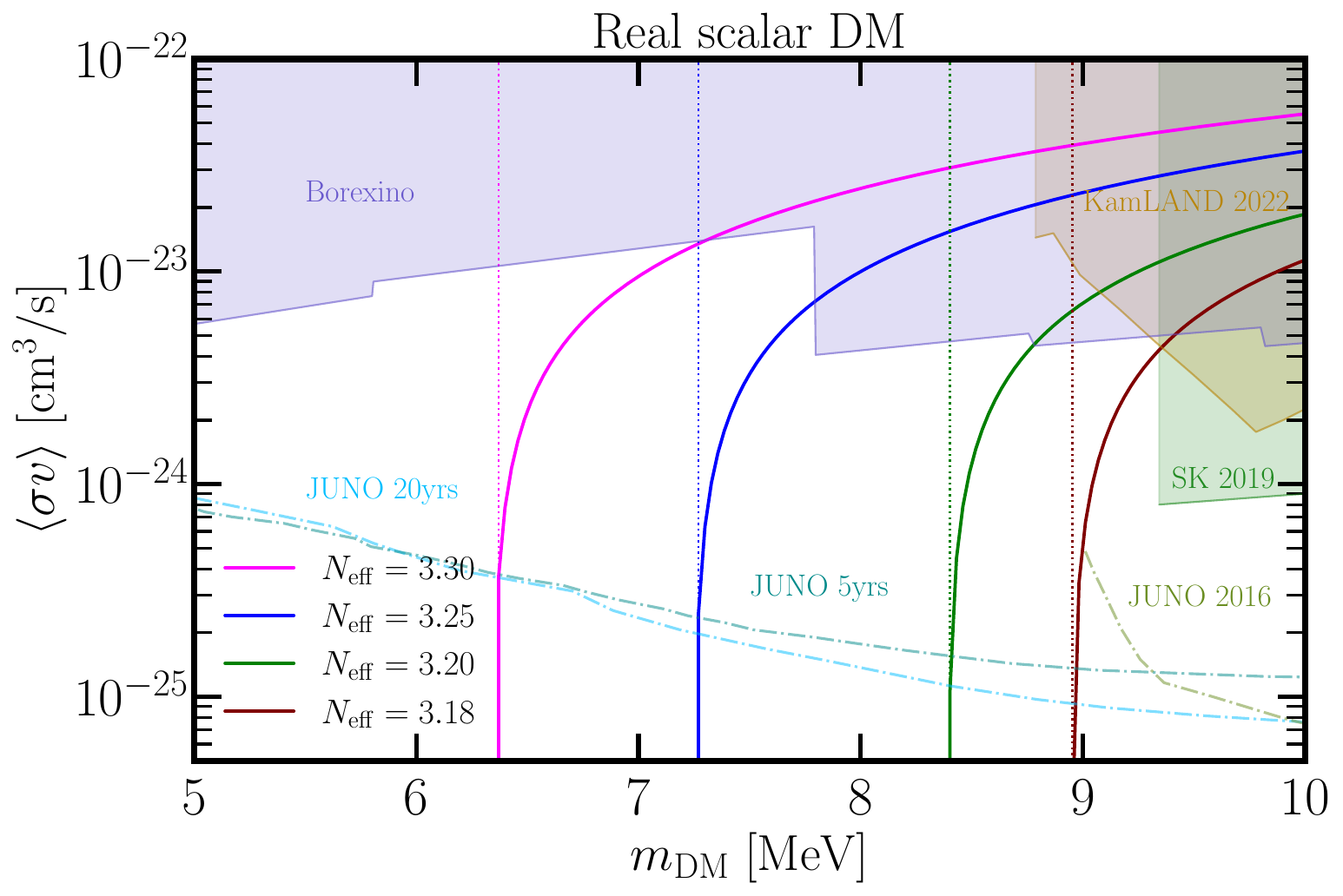} 
	\caption{\label{fig:detection} Limits on thermal real scalar  DM annihilation to neutrinos that predicts $N_{\rm eff}=3.18-3.30$. Also shown are constraints from neutrino detectors and the forecast sensitivities from JUNO.}
\end{figure}

A remarkable  feature appears in  the scalar DM case. We can infer from the left panel of Fig.~\ref{fig:th-bound} that JUNO is expected to probe the parameter space   predicting  $\Delta N_{\rm eff}\simeq 0.2-0.4$. This range of $N_{\rm eff}$ excess, which can be tested with the upgrade of CMB experiments (e.g., CMB-S4 and CMB-HD),  is favored to alleviate the Hubble tension~\cite{DiValentino:2021izs,Schoneberg:2021qvd,Kamionkowski:2022pkx,DiValentino:2025sru}.  We show this feature in Fig.~\ref{fig:detection} by considering $N_{\rm eff}=3.18-3.30$.  We see that the bounds derived from relic DM annihilation are comparable with that from the Borexino, KamLAND and SK experiments. Besides, when the DM mass approaches the lower limit derived by $N_{\rm eff}^{\rm th}$, the bounds get stronger as it should be. This is because of the third  factor mentioned earlier, which points out that  the  room available for relic DM annihilation to generate a large $N_{\rm eff}^{\rm nth}$ becomes strongly suppressed.

In Fig.~\ref{fig:detection}, the forecast sensitivities from JUNO 20yrs and JUNO 2016 are the same as shown in Fig.~\ref{fig:th-bound}, while  we adopt the forecast sensitivity based on 5 years of data-taking  under  a Moore DM profile~\cite{Akita:2022lit}, as shown by  JUNO 5yrs.  With the very recent measurements of   $N_{\rm eff}$, and likely the intriguing  contribution for relaxing the Hubble tension, the JUNO experiments in the upcoming 5-20 years can hopefully provide a complementary test of  the $\Delta N_{\rm eff}\simeq 0.2-0.4$ excess from MeV-scale DM annihilation to neutrinos.  

 In addition to $N_{\rm eff}$,   relic DM annihilation into neutrinos may also  distort the high-energy tail of the cosmic neutrino background (C$\nu$B)~\cite{Dicke:1965zz}, where  neutrino free-streaming can be delayed due to  DM-neutrino scattering.  The theoretical prediction of this effect  depends on the  DM-neutrino  scattering cross section, which generally connects to the DM annihilation cross section  in a model dependent way~\cite{Arguelles:2017atb,Akita:2023yga,Wang:2023csv}.  Nevertheless,  potential  C$\nu$B detection experiments, such as PTOLEMY~\cite{PTOLEMY:2019hkd} and KATRIN~\cite{KATRIN:2022kkv}  capable of observing  C$\nu$B in the eV-MeV range of neutrino energy, will provide another complementary probe of DM-neutrino interaction. 
 
\section{Discussion}

While we have performed model-independent discussions so far, a crucial point  is the realization of a cross section to neutrinos larger than   $\langle \sigma v\rangle_{\rm th}$ and meanwhile the  production of  the  relic density matching the present-day value $\Omega_{\rm DM}h^2\approx 0.12$~\cite{Planck:2018vyg}. A   general scenario that can realize such possibilities is  late-time DM production~\cite{Fairbairn:2008fb, Cheung:2010gj, Medina:2014bga, Gherghetta:2015ysa}. 
Late-time DM production achieves the present-day relic density through mechanisms that produce DM at late times, after thermal freeze-out. A typical example is the   super-WIMP paradigm~\cite{Feng:2003xh,Feng:2003uy}, where the stable DM candidate is produced from long-lived super-WIMP particle decay. In this case,  the DM number density after thermal freeze-out can still be generated in the correct  ballpark, even if most of the thermal relic have been annihilated into neutrinos via a large annihilation cross section. In the  previous model-independent analysis, the  DM number density  
\begin{align}
	n_{\rm DM}&=n_{\rm DM,0}\left(\frac{a_0(t)}{a(t)}\right)^3
	\\[0.2cm]
	&\approx 10^{-12}\left(\frac{T}{0.01~\text{MeV}}\right)^3\left(\frac{1~\text{MeV}}{m_{\rm DM}}\right)~\text{MeV}^3\,,
	\end{align}  
	was used, where $n_{\rm DM,0}$ is the present-day value after redshift with the scale factor $a(t)$. Using  $n_{\rm DM}$ at a given temperature $T$ assumes that  super-WIMP decay is complete prior to $T$, however, the contribution to $N_{\rm eff}$ from  relic DM annihilation is not sensitive to the exact moment of the completion, provided that it occurs after DM/neutrino freeze-out and well before recombination, $T\gg 1$~eV.  In fact, late-time DM production much below 1~keV can suffer from severe constraints from Lyman-$\alpha$~\cite{Irsic:2017ixq,Garzilli:2019qki,Decant:2021mhj}, where DM acquires large momentum and  becomes too warm at present day. Bearing this in mind, we see that a robust bound on relic DM annihilation can still be obtained  largely independent of the exact  production epoch in realistic cases, as inferred from  Eq.~\eqref{eq:Neff-neq-2} that  exhibits a logarithmic dependence on the temperature. 
\section{Conclusion}
The latest measurements of $N_{\rm eff}$ after Planck 2018 have a significant impact on thermal MeV-scale DM that has a large annihilation cross section into neutrinos. They also affect the feasibility of probing such DM particles by observing extra neutrino flux in neutrino detectors. Applying the updated data releases from DESI, SPT-3G and ACT, we have found that the lower mass bound for  MeV-scale DM becomes inconclusive, where significant parameter space can still be targeted by upcoming neutrino experiments. In doing this, we have  included the relic DM annihilation effect on late-time contributions to $N_{\rm eff}$. This treatment strengthens the upper bounds of the annihilation cross section to neutrinos, which are comparable with or stronger than the current bounds from neutrino detectors.  We have also illustrated the complementary probes of real scalar DM, by showing that    an excess of $N_{\rm eff}$ at $0.2-0.4$ remarkably resides in the region  that can be  probed by upcoming JUNO experiments.

\section*{Acknowledgements}
This work is supported by JSPS Grant-in-Aid for JSPS Research Fellows No.~24KF0060 (S.K. and S.-P.Li) and No.~24KF0238 (S.K. and D.N.). S.K. is also supported in part by Grants-in-Aid for Scientific Research(KAKENHI) No.~23K17691 and No.~20H00160.


\bibliographystyle{utphys}
\bibliography{Refs}

\end{document}